\documentclass[12pt]{article}
\usepackage{setspace}
\usepackage{amssymb,latexsym,amsmath} 

\usepackage{epsf,subfigure}

\author{Lauren M. Childs and Steven H. Strogatz\\
Center for Applied Mathematics, \\  Cornell University, Ithaca, NY 14853 USA \\
\\
\texttt{lmchilds@cam.cornell.edu, strogatz@cornell.edu}
\bigskip
\bigskip
\bigskip
}

\title{Stability diagram for the forced Kuramoto model}

\begin{document}

\maketitle

\abstract{We analyze the periodically forced Kuramoto model.  This system consists of an infinite population of phase oscillators with random intrinsic frequencies, global sinusoidal coupling, and external sinusoidal forcing. It represents an idealization of many phenomena in physics, chemistry and biology in which mutual synchronization competes with forced synchronization.  In other words, the oscillators in the population try to synchronize with one another while also trying to lock onto an external drive.  Previous work on the forced Kuramoto model uncovered two main types of attractors, called forced entrainment and mutual entrainment, but the details of the bifurcations between them were unclear.  Here we present a complete bifurcation analysis of the model for a special case in which the infinite-dimensional dynamics collapse to a two-dimensional system.  Exact results are obtained for the locations of Hopf, saddle-node, and Takens-Bogdanov bifurcations.  The resulting stability diagram bears a striking resemblance to that for the weakly nonlinear forced van der Pol oscillator.  }

\bigskip
\bigskip
\bigskip

\noindent Abbreviated title: Forced Kuramoto model 

\pagestyle{headings}

\newpage

\noindent \textbf{The study of synchronization is a classic topic in nonlinear science. Sometimes the concern is with mutual synchronization, as in Huygens's 1665 discovery of the sympathy of pendulum clocks.  In other situations, one is more interested in forced synchronization, as in the injection locking of a laser or the entrainment of circadian rhythms by the daily light-dark cycle.  Here we consider a simple mathematical model in which both types of synchronization are present simultaneously, creating a conflict between them.  What happens when a network of dissimilar but mutually coupled oscillators is also driven by an external periodic force?  For a natural generalization of the Kuramoto model, the interaction of forcing, coupling, and randomness leads to a rich set of collective states and bifurcations.  We explain all of these phenomena analytically, using an ansatz recently introduced by Ott and Antonsen.} 

\section{Introduction}

In 1975 Kuramoto proposed an elegant model for an enormous population of coupled biological oscillators [Kuramoto 1975, 1984].  Each oscillator was described solely by its phase, with amplitude variations neglected; the oscillators were coupled all-to-all, with equal strength; the interaction between them was purely sinusoidal, with no higher harmonics; and their intrinsic frequencies were randomly distributed across the population according to a symmetric bell-shaped distribution.  All of these simplifying assumptions helped Kuramoto make headway on what would otherwise have been a hopelessly intractable many-body, nonlinear dynamical system.  By means of an ingenious self-consistency argument, he was able to show analytically that the system could undergo a phase transition to mutual synchronization, once the coupling between the oscillators exceeded a certain threshold.  

Over the past three decades, many researchers have shed light on the mathematical aspects of collective synchronization by studying Kuramoto's model and its close relatives [Strogatz 2000, Acebron et al. 2005].  And, somewhat surprisingly in view of its simplicity, the model has also been shown to be relevant to a variety of physical systems [Pikovsky et al. 2001, Strogatz 2003].  Examples range from electrochemical oscillators [Kiss et al. 2002, 2008] and Josephson junction arrays [Wiesenfeld et al. 1996] to coupled metronomes [Pantaleone 2002], collective atomic recoil lasing [von Cube et al. 2004],  and neutrino flavor oscillations [Pantaleone 1998].

One way to extend the model is to allow for the effects of external forcing.  This generalization is theoretically natural, but it is also motivated in part by experimentally observed phenomena [Kiss et al. 2008].  For example, consider the way that the daily cycle of light and darkness helps to entrain our sleep, body temperature, and other circadian rhythms to the world around us [Moore-Ede et al. 1982, Dunlap et al. 2003, Foster and Kreitzman 2005].  Like all mammals, each of us has a circadian pacemaker, a network of thousands of specialized clock cells located in the region of the hypothalamus known as the suprachiasmatic nuclei, just above where the optic nerves criss-cross as they make their way back to the brain.  These cells have been shown experimentally to be intrinsically oscillatory [Welsh et al. 1995] and their distribution of natural frequencies has been measured [Liu et al. 1997].  The pacemaker cells are also known to be mutually coupled, though their precise connectivity remains unclear.  Thus, qualitatively at least, one could try to model the pacemaker cell network with the Kuramoto model.  Now consider how this network might respond to an imposed cycle of light and dark (information of this sort is known to be conveyed from the eyes to the pacemaker through a specialized neural pathway).  If the light-dark cycle is 24 hours long, we expect the electrical rhythms of many individual pacemaker cells to successfully entrain to it.   But what if we alter the period or strength of the external forcing, as has been done in countless experiments on mice, hamsters, primates, and human volunteers [Moore-Ede et al. 1982]?  Or what happens if the experiment is conducted on mutant organisms [Dunlap 1993, Takahashi 1995] whose intrinsic periods are a few hours longer or shorter than normal, or which may be intrinsically arrhythmic, having almost no free-running circadian rhythm at all?  

Questions like this can be addressed, in mathematically idealized form, within the framework of the periodically forced Kuramoto model [Sakaguchi 1988, Antonsen et al. 2008, Ott and Antonsen 2008].  Its governing equations are given by

\begin{equation}\label{forcedKuramoto_labframe}
\frac{d \vartheta_i}{dt} = \omega_i+\frac{K}{N}\sum_{j=1}^N \sin(\vartheta_j-\vartheta_i)+F \sin(\sigma  t-\vartheta_i),
\end{equation}

\noindent for $i = 1, \ldots, N$.  Here $\vartheta_i$ is the phase of oscillator $i$, $\omega_i$ is its natural frequency, $K$ is the coupling strength, $F$ is the forcing strength, $\sigma$ is the forcing frequency, and $N \gg 1$ is the number of oscillators.  The natural frequencies are randomly distributed with a density $g(\omega)$, assumed unimodal and symmetric about its mean value $\omega_0$.  

This system is capable of rich dynamics because of its interplay among randomness, coupling, and forcing.  The randomness comes from the variance of the natural frequencies.  This effect tends to desynchronize the oscillators and scatter their phases.  The coupling, on the other hand, tends to align the oscillators to the same phase, although it does not  favor any particular frequency for the collective oscillation.  In contrast, the forcing does favor a specific frequency, namely that of the external drive.  Depending on the relative magnitudes of these competing effects, we expect to see various kinds of cooperative behavior and transitions between them.

Before continuing, it proves useful to simplify the governing equations in two ways.  First, if we view the dynamics in a frame co-rotating with the drive, the explicit time dependence in \eqref{forcedKuramoto_labframe} disappears.   To achieve this, let 
\begin{equation}\label{rotatingframe}
\theta_i = \vartheta_i - \sigma  t.
\end{equation}
Then \eqref{forcedKuramoto_labframe} yields 
\begin{equation}\label{forcedKuramoto}
\frac{d \theta_i}{dt} = (\omega_i - \sigma) +\frac{K}{N}\sum_{j=1}^N \sin(\theta_j-\theta_i) - F \sin \theta_i,
\end{equation}
\noindent for $i = 1, \ldots, N$.  Second, as Kuramoto originally pointed out, it is helpful to introduce a complex order parameter $z$, given by 
\begin{equation}\label{complexorderparameter}
z(t)=\frac{1}{N} \sum_{j=1}^N e^{i \theta_j(t)}.
\end{equation}  
Then the sum in \eqref{forcedKuramoto} reduces to \textrm{Im}$(K z e^{-i \theta_i})$, an identity which will prove useful later.  

The order parameter also has a nice physical interpretation.  Its amplitude $|z|$ quantifies the phase coherence of the population: an incoherent state has $z = 0$; a perfectly coherent state has $|z| = 1$.  Furthermore, the argument of $z$ can be interpreted as the average phase of all the oscillators.  So in a sense, the single complex number $z(t)$ serves as a proxy for the state of the population as a whole.  

Sakaguchi [1988] was the first to study the periodically forced Kuramoto model.  He derived a self-consistent equation for steady-state values of $|z|$, under the assumption that $z(t)$ was entrained by the external force (meaning that $z(t)$ appeared motionless in the rotating frame).  In numerical simulations of Eq.~\eqref{forcedKuramoto}, however, Sakaguchi found that this state of ``forced entrainment'' was not always attained.  For some values of the parameters, the system could settle instead into a state of ``mutual entrainment.''  In this case a macroscopic fraction of the system self-synchronized at a different frequency from that of the drive, indicating that this sub-population had broken away and established its own collective rhythm.  (For circadian rhythms, this would mean that the animal's internal clock was drifting relative to the outside world.)  Sakaguchi's numerics further indicated how forced entrainment could be lost and give way to mutual entrainment.  Such transitions were found to occur via two different mechanisms, corresponding to a pair of distinct bifurcation curves  in parameter space.  These curves appeared to join at a point, but Sakaguchi was unable to resolve the details of the cross-over region numerically.

More recently, Antonsen et al. [2008] gave an improved analytical treatment of the model.  Their linear stability analysis and numerical simulations also revealed an intriguing set of bifurcation curves, but the way the various curves join together still remained unclear.  The overall layout of the stability diagram suggested that an underlying two-dimensional system was controlling the dynamics---a remarkable finding, given that the model~\eqref{forcedKuramoto} is essentially infinite-dimensional (recall $N \gg 1$).   

This tantalizing clue led Ott and Antonsen to an important discovery [Ott and Antonsen 2008].  They found that the Kuramoto model possesses an invariant manifold, a special family of states for which the macroscopic dynamics becomes \emph{low}-dimensional.  In particular they showed that on this invariant manifold, the order parameter for the forced Kuramoto model~\eqref{forcedKuramoto} exactly satisfies a \emph{two}-dimensional dynamical system, for the special case where the frequency distribution $g(\omega)$ is Lorentzian and the initial state satisfies certain strong analyticity properties with respect to $\omega$.

In this paper we analyze the two-dimensional system derived from the analysis of Ott and Antonsen [2008].  Our results give the first complete picture of the bifurcation structure for the forced Kuramoto model.   We obtain explicit formulas for the system's saddle-node and Hopf bifurcation curves, as well as the codimension-2 Takens-Bogdanov point from which they emanate.  Bifurcation theory predicts that a curve of homoclinic bifurcations should also emerge from the Takens-Bogdanov point; we compute this homoclinic curve numerically.  

The rest of the paper is organized as follows.  Section 2 reviews the approach of Ott and Antonsen [2008], leading up to their derivation of the reduced equations for the order parameter dynamics.  Section 3 presents new results about the bifurcations in this system and resolves the issue of how all the transition curves fit together.  The final section discusses the implications of the results, their relation to prior work, the limitations of the approach used here, and some of the questions that remain.  

\section{Derivation of the reduced equations}

The analysis of \eqref{forcedKuramoto} is carried out in the continuum limit $N \rightarrow \infty$.  Then the state of the system is described by a density function $f(\theta, \omega, t)$.  Here $f$ is defined such that at time $t$, the fraction of oscillators with phases between $\theta$ and $\theta + d\theta$ and natural frequencies between $\omega$ and $\omega + d \omega$ is given by $f(\theta, \omega, t) \, d\theta \, d\omega$.  Thus 
\begin{equation}\label{normalization}
 \int_{-\infty}^\infty \int_0^{2 \pi} f(\theta, \omega,t)  \, d \theta \,d \omega  = 1
\end{equation}
and
\begin{equation}\label{freqdist}
\int_0^{2 \pi} f(\omega,\theta,t) \, d \theta = g(\omega),
\end{equation}
by definition of $g(\omega)$.  

The evolution of $f$ is given by the continuity equation 
\begin{equation}\label{continuitysimple}
\frac{\partial f}{\partial t} + \frac{\partial}{\partial \theta} (f v) = 0,
\end{equation}
which expresses the conservation of oscillators of frequency $\omega$.  Here $v(\theta, \omega, t)$ is the velocity field on the circle corresponding to \eqref{forcedKuramoto} as $N \rightarrow \infty$:
\begin{equation}\label{velocity}
v(\theta, \omega, t) = (\omega-\sigma) + K \int_{-\infty}^\infty  \int_0^{2 \pi} \sin(\theta'-\theta) \, f(\theta', \omega',t) \, d \theta'  \,d \omega' - F \sin \theta.
\end{equation}
This expression can be written more compactly in terms of the complex order parameter $z$, which in the continuum limit becomes 
\begin{equation}\label{z_continuum}
z(t) =  \int_{-\infty}^\infty \int_0^{2 \pi} e^{i \theta} f(\theta, \omega,t)  \,d \theta \,d \omega.
\end{equation}
Using the identity mentioned in the Introduction, we note that the double integral in \eqref{velocity}  simplifies to \textrm{Im}$(K z e^{-i \theta})$.  Hence the continuity equation becomes 
\begin{equation}\label{Cont}
\frac{\partial f}{\partial t} + \frac{\partial}{\partial \theta}\left( f \left[ (\omega-\sigma) + \frac{1}{2 i} \left\{ (K z + F) e^{-i \theta}-(K z + F)^* e^{i \theta} \right\} \right] \right) = 0,
\end{equation}
where the asterisk denotes complex conjugation.  

Normally one would try to solve  \eqref{Cont} by expanding $f$ as a Fourier series in $\theta$:
\begin{equation}\label{fourier}
f(\theta, \omega, t) = \frac{g(\omega)}{ 2 \pi} \left[ 1 + \sum_{n=1}^\infty f_n(\omega, t) e^{i n \theta} + \rm{c.c.} \right],
\end{equation}
where c.c. denotes complex conjugate.  Substitution of \eqref{fourier} into \eqref{z_continuum} and \eqref{Cont}  would generate an infinite set of coupled nonlinear ordinary differential equations for the amplitudes $f_n(\omega, t)$.  Unfortunately the dynamics of this infinite-dimensional system would likely be difficult to analyze further.  

It was at this point that Ott and Antonsen [2008] noticed something wonderful.  They restricted attention to the special family of densities $f$ for which 
\begin{equation}\label{trick}
f_n(\omega, t) = \left[\alpha(\omega, t) \right]^n, 
\end{equation}
for all $n \ge 1$.  In other words, they assumed that all the amplitudes $f_n$ are $n^{\rm th}$ powers of the \emph{same} function $\alpha(\omega, t)$.  Amazingly, this ansatz satisfies the amplitude equations for all $n$, so long as $\alpha$ evolves according to 
\begin{equation}\label{alphaeqn}
\frac{d \alpha}{d t} = \frac{1}{2} (K z + F)^{*} - i (\omega - \sigma) \alpha - \frac{1}{2} (K z + F) \alpha^2
\end{equation}
and $z$ satisfies
\begin{equation}\label{zintegral}
z(t) = \int_{-\infty}^{\infty} \alpha^{*}(\omega, t) \, g(\omega) \, d \omega .
\end{equation}
Then, by further assuming that  $g(\omega)$ is a Lorentzian,
\begin{equation}\label{lorentzian}
g(\omega)=\frac{\Delta}{\pi \left\{ (\omega-\omega_0)^2+\Delta^2 \right\}},
\end{equation}
and that $\alpha(\omega, t)$ satisfies certain analyticity conditions in the complex $\omega$-plane, Ott and Antonsen [2008] evaluated \eqref{zintegral} by contour integration and thereby derived the following exact evolution equation for the order parameter $z$:
\begin{equation}\label{orderparamtimeevol}
\frac{dz}{dt} = \frac{1}{2}\left[ (K z + F) - (K z + F)^* z^2  \right] - \left[ \Delta +i (\sigma - \omega_0)\right] z.
\end{equation}
The conditions required were that $\alpha(\omega, t)$ can be analytically continued from the real $\omega$-axis into the lower half of the complex $\omega$-plane for all $t \ge 0$; that $|\alpha(\omega, t)| \rightarrow 0$ as $\rm{Im}(\omega) \rightarrow -\infty$; and that $|\alpha(\omega, 0)| \le 1$ for real $\omega$.  

\section{Analysis of the reduced equations} 
 
\subsection{Scaling the equations}
We turn now to the analysis of the two-dimensional system \eqref{orderparamtimeevol}. The first step is to reduce the number of parameters by nondimensionalizing the system.   Let $\hat t = \Delta t, \hat F = F/\Delta, \hat K = K/\Delta, \hat \sigma = \sigma/\Delta$ and $ \hat \omega_0 = \omega_0/\Delta$.  Then the form of \eqref{orderparamtimeevol} stays the same except that $\Delta$ no longer appears (in effect, $\Delta$ has been set to 1 without loss of generality) and all the other parameters now have hats over them.  For ease of notation we drop the hats in what follows, remembering that all the parameters are now dimensionless.  Also, let 
\begin{equation}\label{detuning}
\Omega = \sigma - \omega_0
\end{equation}
denote the (dimensionless) detuning between the drive frequency and the population's mean natural frequency.   Then if we introduce polar coordinates  
\begin{equation}\label{polar}
z = \rho e^{i \phi}
\end{equation}
and separate \eqref{orderparamtimeevol} into real and imaginary parts, we obtain the dimensionless evolution equations for $\rho$ and $\phi$:

\begin{eqnarray}
\rho' &=& \frac{K}{2} \rho (1 - \rho^2) - \rho + \frac{F}{2}(1- \rho^2) \cos{\phi} \label{eq:phi-rhoa}\\
\phi' &=& -\left[\Omega + \frac{F}{2}\left(\rho + \frac{1}{\rho}\right) \sin{\phi}\right] \label{eq:phi-rhob}
\end{eqnarray}
where the prime denotes differentiation with respect to dimensionless time.

\subsection{Stability diagram and phase portraits}

Our next goal is to obtain the stability diagram for Eqs.~(\ref{eq:phi-rhoa})-(\ref{eq:phi-rhob}).  Before delving into the details, which can become intricate at times, we jump to the final result.  Figure 1 shows the stability diagram for the representative case where $K=5$.  Here the various stability regions labeled A-E correspond to the phase portraits shown in Fig.~2.  

We realize that these figures appear complicated at first glance, so let us begin by offering a few general remarks about them.  Figure 1 is divided into five regions, A-E, by the bifurcation curves labeled saddle-node, Hopf, homoclinic, and SNIPER.  In the places where two or more of these curves nearly coincide, Fig.~1(a) becomes especially confusing.  To clarify what is going on in such regions, Figs.~1(b) and 1(c) zoom in near two codimension-2 points of interest (to be discussed in detail later).  Since even these figures can be hard to interpret, we have tried to make everything as clear as possible by presenting a schematic Fig.~1(d).  Unlike Figs.~1(a)-(c), which are numerically accurate, Fig.~1(d) is only topologically correct.  We have distorted some of stability regions and pulled certain curves apart to make the layout of the diagram transparent, and to highlight the three different codimension-2 points that will later be seen to organize the entire diagram.  
 
A similar but incomplete version of Fig.~1 was obtained previously by Antonsen et al. [2008]; see Fig.~3 in their paper.  Those authors generated their results based on direct simulations of Eq. \eqref{forcedKuramoto} for $N=1000$ oscillators.  They also compared their numerics to analytical results they derived for the existence and stability of equilibrium points for \eqref{forcedKuramoto}, which correspond to entrained states in the original frame.  Our approach, in contrast, is to analyze the reduced system Eqs.~(\ref{eq:phi-rhoa})-(\ref{eq:phi-rhob}).  We do not present numerical results for the full system \eqref{forcedKuramoto} because in every case we have checked, our results match those reported already by Antonsen et al. [2008], except in cases where the previous methods were inconclusive.  

\subsection{Saddle-node and SNIPER bifurcations}
It is algebraically awkward to solve for the fixed points of Eqs.~(\ref{eq:phi-rhoa})-(\ref{eq:phi-rhob}) in terms of the parameters.  Fortunately, we do not need to solve for them.  Since we are mainly interested in the bifurcation curves, we can make headway more easily by imposing an appropriate bifurcation condition and then solving for the parameters in terms of the fixed point, rather than the other way around.  This is a standard trick in bifurcation theory, and it allows us to derive the bifurcation curves in closed form, either explicitly or as parametric equations.  

For example, at a saddle-node bifurcation, one of the eigenvalues equals 0 and hence the determinant of the Jacobian vanishes there.  (The same would be true at transcritical or pitchfork bifurcations, but given the absence of the constraints or symmetries associated with these types of bifurcations, there's no reason to expect either of them to occur here.)  

Hence to find the locus of saddle-node bifurcations, we solve $\rho' = 0$, $\phi' = 0$ and $\delta=0$ simultaneously, where $\delta$ denotes the determinant of the Jacobian.  The trick is to regard the unknown values of the variables $\rho$ and $\phi$ on equal footing with the parameters $K, \Omega$ and $F$.  Then the resulting system of 3 equations in 5 unknowns can be solved explicitly to yield a parametrization of the saddle-node bifurcation surface.  Various parametrizations are possible.  One convenient choice is to express the parameters in terms of the fixed-point values of $\rho$ and $\phi$.  We find that the saddle-node surface is then given by
\begin{eqnarray}\label{SNsurface}
K&=&\frac{2 \left(\rho ^4+2 \rho ^2 \cos 2 \phi  +1\right)}{\left(1-\rho
   ^2 \right)^2 \left(1+ \rho ^2 \cos 2 \phi \right)}\\
\Omega &=&\frac{\left(\rho ^3+\rho \right)^2 \sin 2 \phi }{\left(1-\rho
   ^2 \right)^2 \left(1+ \rho ^2 \cos 2 \phi \right)}\\
F&=&-\frac{4 \rho ^3 \left(\rho ^2+1\right) \cos \phi }{\left(1-\rho
   ^2 \right)^2 \left(1+ \rho ^2 \cos 2 \phi \right)}
\end{eqnarray}
where we allow $\rho$ and $\phi$ to sweep over their full ranges $0 \le \rho \le1$, $-\pi \le \phi \le \pi$.  

This parametrization provides some interesting information.  For instance, it shows that $K$ increases monotonically with $\rho$, for each fixed value of $\phi$. Hence $K \ge  2$ everywhere on the saddle-node surface, with the minimum value $K=2$ being attained when $\rho=0$ and hence $F=0$, or in other words, when there is no forcing.  This result makes sense.  In the absence of forcing, the system is just the traditional Kuramoto model with a Lorentzian $g(\omega)$, and $K = 2 \Delta$ (or in dimensionless terms, $K=2$) is the well-known formula for the critical coupling at the onset of mutual synchronization [Kuramoto 1975, 1984].  

To compare our results with those obtained numerically by Antonsen et al. [2008], it is more illuminating to slice through the saddle-node surface at a fixed value of $K>2$ and then plot the resulting saddle-node curves in the $(\Omega, F)$ plane.  To find these curves we solve $\rho' = 0$, $\phi' = 0$ and $\delta=0$ for $\Omega, \sin \phi$ and $\cos \phi$, and then use $\sin^2 \phi+ \cos^2 \phi = 1$ to solve for $F$, now regarding $K$ and $\rho$ as parameters.  The result is the following parametrization of the saddle-node curve:
\begin{eqnarray}\label{SNcurvesK}
\Omega_{\rm \,SN} &=& \frac{\left(\rho ^2+1\right)^{3/2}
   \sqrt{K \left(\rho ^2-1\right)
   \left(K \left(\rho
   ^2-1\right)^2-4\right)-4
   \left(\rho ^2+1\right)}}{2
   \left(\rho ^2-1\right)^2}\\
F_{\rm \,SN} &=& \frac{\sqrt{2} \rho ^2 \sqrt{K^2
   \left(\rho ^2-1\right)^3+2 K
   \left(\rho ^4-4 \rho
   ^2+3\right)-8}}{\left(\rho
   ^2-1\right)^2}.
\end{eqnarray}
Figure~1 plots this saddle-node curve for the case $K=5$, as previously studied by Antonsen et al. [2008].  We compute the curve for all values $0<\rho  <1$, disregarding any values that yield non-real results for $\Omega$ or $F$.  

The two branches of the saddle-node curve intersect tangentially at a codimension-2 cusp point, as highlighted in Fig.~1(c) and marked schematically in  Fig.~1(d) by the solid square .  For $K=5$, the parameter values at the cusp are $\Omega \approx 3.5445$ and $F \approx 3.4164$.  

Along with local saddle-node bifurcations, the lower branch of the saddle-node curve(where $F \approx \Omega$) also includes a large section consisting of saddle-node infinite-period (SNIPER) bifurcations.   These have important global implications, because they create or destroy limit cycles in the phase portrait of Eqs.~(\ref{eq:phi-rhoa})-(\ref{eq:phi-rhob}).   

\subsection{Hopf bifurcation}

Next we calculate the locus of parameter values at which Hopf bifurcations occur.  We impose the fixed point conditions $\phi' = 0$, $\rho' = 0$ as before, but now require that the Jacobian has zero trace and positive determinant---the latter two conditions are equivalent to requiring that the eigenvalues be pure imaginary.

Solving simultaneously for $\phi' = 0$, $\rho' = 0$ and trace = 0, we find 
\begin{eqnarray}\label{zerotrace_conditions}
\sin \phi
   &=& -\frac{\left(K^2-4\right)
   \Omega }{F \sqrt{K-2} K
   \sqrt{K+2}}\\
\cos \phi &=& -\frac{(K-2)^{3/2}}{2
   F \sqrt{K+2}}\\
\rho &=&\sqrt{\frac{K-2}{K+2}}.
\end{eqnarray}

Because $\rho$ depends only on $K$, we can go a bit further than we did in the saddle-node case.  Using $\sin^2 \phi+ \cos^2 \phi = 1$ as before,  $F$ can now be obtained \emph{explicitly} in terms of $K$ and $\Omega$:
\begin{equation}\label{HopfcurvesK}
F_{\rm \,Hopf}=\frac{1}{2K} \sqrt{\frac{(K-2)
   \left(K^4-4 K^3+4 \left(\Omega
   ^2+1\right) K^2+16 \Omega ^2
   K+16 \Omega ^2\right)}{K+2}}
\end{equation}
For the special case $K=5$ studied by Antonsen et al. [2008], Eq. \eqref{HopfcurvesK} becomes 
\begin{equation}
F_{\rm \,Hopf} = \frac{1}{10} \sqrt{\frac{3}{7}} \sqrt{225+ 196 \Omega^2}
\end{equation}
Figure 1 plots the graph of $F_{\rm \,Hopf}(\Omega)$.  Notice how straight it is, and that it nearly lines up with the lower branch of the saddle-node curve.   

\subsection{Takens-Bogdanov point}

As mentioned above, for Eq.\eqref{HopfcurvesK} to truly signify a Hopf bifurcation, the Jacobian determinant must be positive at the corresponding parameter values $(\Omega, F)$ .  This will be the case if $\Omega$ and $F$ are sufficiently large.  Specifically, their values must exceed those at the Takens-Bogdanov point 
\begin{eqnarray}
\Omega_{\rm \,TB} &=& \frac{(K-2) K^2}{4 (K+2)} \label{TB_Ka}\\
F_{\rm \,TB} &=& \frac{1}{4} (K-2)
   \sqrt{\frac{K^3-2 K^2+4
   K-8}{K+2}} \label{TB_Kb}
\end{eqnarray}
obtained by solving four simultaneous equations: $\phi' = 0$, $\rho' = 0$, trace = 0, and determinant = 0.  

The Takens-Boganov point is marked with a filled circle on Figs.~1(a) and 1(d). In addition to serving as the endpoint of the Hopf curve, it splits the upper branch of the saddle-node curve into two sections of different dynamical character.  On the lower section (below the Takens-Boganov point), an \emph{unstable} node collides with a saddle along the saddle-node bifurcation curve; this can be seen by comparing regions D and A, as shown in Figs.~2(d) and 2(a).  The opposite situation occurs on the upper section of the saddle-node curve, above the Takens-Boganov point.  Here a \emph{stable} node collides with a saddle, corresponding to the transition between regions B and A; see Figs.~2(b) and 2(a).

\subsection{Homoclinic bifurcation}

The theory of the Takens-Bogdanov bifurcation implies that a curve of homoclinic bifurcations must also emerge from the codimension-2 point, tangential to the saddle-node and Hopf curves.  For the case $K=5$ shown in Fig.~1, $\Omega_{\rm \,TB} = \frac{75}{28}$ and $F_{\rm \,TB} = \frac{3}{4}\sqrt{\frac{87}{7}}$.  The curve shown in the diagram was computed numerically. It is almost indistinguishable from the Hopf curve and thus produces a very small region between them, as shown in Fig.~1(b).  

A striking feature of the homoclinic curve is that after moving parallel to the Hopf curve for a while, it makes a sharp backward turn and then joins onto the lower branch of the saddle-node/SNIPER curve, meeting that curve tangentially at a codimension-2 ``saddle-node-loop'' point [Guckenheimer 1986, Izhikevich 2000] marked by a filled diamond in Figs.~1(b) and 1(d).

\subsection{Phase portraits and bifurcation scenarios}

As we have seen, the bifurcation curves in Fig.~1 partition the stability diagram into five regions, labeled A-E.  We now give a fuller treatment of the dynamics associated with each region and the transitions from one to another.

\subsubsection{Region A: Forced entrainment}  
Here the order parameter $z$ approaches a stable fixed point for all initial conditions, as shown in Fig. 2(a).  To interpret what this means, recall that all our analysis has assumed a frame co-rotating with the drive.  Hence this stable fixed point represents a state in which the order parameter is moving periodically while staying phase-locked to the drive.  Therefore, back in the original frame, a macroscopic fraction of the oscillator population must also be moving in rigid synchrony, locked to the same frequency as the drive signal.    

\subsubsection{Region B: Bistability between two states of forced entrainment}  
Now suppose we weaken the forcing.  Imagine moving down along a vertical line in Fig.~1(b), decreasing $F$ while holding $\Omega$ fixed.   As we do so, we first pass from region A into the extremely narrow region B by crossing through the upper branch of the saddle-node curve \eqref{SNcurvesK}.  At this bifurcation, a stable node is born out of the vacuum, along with a saddle point.  Meanwhile, the stable fixed point that we encountered in Region A still exists; it lies in the lower right part of Fig.~2(b).  

Thus Region B depicts a form of bistability.  Depending on the initial conditions, the system chooses one of two possible states of forced entrainment, differing in their phase coherence (the magnitude of $z$) and their phase relationship to the drive signal (the argument of $z$).  

\subsubsection{Region C: Bistability between forced entrainment and phase trapping}  
Continuing our vertical descent through Fig.~1(b), we next cross from B into C by passing through the curve of Hopf bifurcations,  Eq.~\eqref{HopfcurvesK}.  The stable fixed point created in Region B now loses stability and gives birth to a tiny attracting limit cycle (Fig.~2(c)).  On this cycle the order parameter still runs at the same average frequency as the drive but its relative phase and amplitude now wobble slightly.  Because these variations remain trapped within tight limits, one says the system is phase trapped (as opposed to phase locked) to the drive.   Back in the original non-rotating frame, the macroscopic dynamics for this state would be quasiperiodic with two frequencies.  This is not the only attractor, of course; the state of forced entrainment seen earlier in A and B persists, so we still have bistability, but now between phase trapping and forced entrainment.

\subsubsection{Region D: Forced entrainment}
Passing from Region C to D carries us across a curve of homoclinic bifurcations.  As we approach this curve from above, the tiny limit cycle in Fig.~2(c) expands.  At the bifurcation it touches the saddle point and forms a homoclinic orbit.  Beyond the bifurcation the phase portrait looks like that shown in Fig.~2(d).  An invariant loop has been created, in which the saddle and the original stable node are now connected by both branches of the saddle's unstable manifold.  The stable node is the unique attractor.  Hence the system again falls into a state of forced entrainment.  

\subsubsection{Region E: Mutual entrainment}
Forced entrainment is finally lost when we pass from Region D to E.  When crossing the lower branch of the saddle-node curve, we need to be careful to specify exactly where the crossing occurs.  Specifically, do we cross to the left or right of the codimension-2 saddle-node-loop point  (filled diamond in Fig.~1(b)) at which the homoclinic curve meets the saddle-node curve?  

Suppose first that we cross below and to the left of the saddle-node-loop point.  Then in Fig.~2(d) the saddle and node would slide toward each other along the invariant loop, coalesce, and disappear, leaving a stable limit cycle in their wake.  Thus, this saddle-node bifurcation is actually a SNIPER (saddle-node infinite-period) bifurcation.  

The limit cycle created by the bifurcation is globally attracting.  Hence the order parameter always settles into periodic motion in the rotating frame.  But unlike the limit cycle of Fig.~2(b) this cycle winds around the origin of the $z$-plane, marked by an asterisk in Fig.~2(d).  This is an important distinction, because it implies that the phase of $z$ now increases monotonically relative to that of the drive. Consequently the order parameter $z(t)$ oscillates at a different average frequency from the drive signal, implying that a macroscopic fraction of the oscillator population has broken loose from the drive.  In other words, the system has spontaneously mutually entrained itself, at least in part.  This is therefore one mechanism by which forced entrainment can give way to mutual entrainment.  

But there are other possible mechanisms as well.  For example, consider Fig.~1(b) again, and now direct your attention to the upper right corner.  By moving down along the right side of the picture, we can cross directly from C to E, without ever going through D.  This happens when we cross through the portion of the lower saddle-node curve lying above and to the right of the saddle-node-loop point. In this case the bifurcation is not a SNIPER; it's just an ordinary saddle-node bifurcation. To visualize this scenario, imagine sliding the saddle in the middle of Fig.~2(c) to the right along its unstable manifold until it collides with the node and annihilates it.  During this process the limit cycle in Fig.~2(c) grows.  And so the phase portrait now resembles the one shown in Fig.~2(e).

A third scenario is much simpler.  Suppose $\Omega > \Omega_{\rm \, cusp}$, so that we're well to the right of the cusp in Figs.~1(c) and 1(d).  Then as we decrease $F$, we move directly from A to E.  Forced entrainment gives way to mutual entrainment through a supercritical Hopf bifurcation.

\section{Discussion}

\subsection{Stability diagram}
The main result of the paper is the stability diagram shown in Fig.~1.  We have focused on the analytical derivation of several of the curves in this picture and tried to clarify how they fit together and what they imply about the system's overall dynamics.  Having immersed ourselves in the details, it is worthwhile to step back and try to understand the broader lessons that this picture holds.

Figure 1 essentially divides into two big regions.  One represents forced entrainment, wherein a macroscopic fraction of the population is phase-locked to the drive.  The rest of the population consists of oscillators belonging to the infinite tails of the frequency distribution; these remain unlocked.  Thus it would be more accurate to speak of ``partial forced entrainment,'' though we hope the intended meaning of the shorter name is clear.

The other main region represents (partial) mutual entrainment.  Now there are three qualitatively different groups of oscillators: (1) the unlocked oscillators in the tails; (2) the oscillators entrained by the forcing; and (3) a self-organizing group of oscillators that entrain one another at a frequency different from that of the drive. The existence of this third group causes the order parameter to wobble or drift periodically relative to the drive, as manifested by a stable limit cycle in the phase portraits (Figs.~2(c) and 2(e)).  

\subsection{Comparison to Adler equation}
The boundary between forced and mutual entrainment is complicated when viewed at a fine scale, as shown in Fig.~1(b).  But from a bird's-eye view, it looks very much like the straight line $F = \Omega$.  Here's why: this is the result one would expect from the Adler equation 
\begin{equation}\label{adler}
 \phi' = -\Omega - F \sin \phi,
\end{equation}
which has been used to model the entrainment dynamics of phase-locked loops [Adler 1946], lasers [Siegman 1986, Yeung and Strogatz 1998], and fireflies [Ermentrout and Rinzel 1984], among many other systems.  The two-dimensional system~(\ref{eq:phi-rhoa})-(\ref{eq:phi-rhob}) reduces to Adler's equation as $K \rightarrow \infty$, in the sense that $\rho$ approaches 1 on a fast time scale, while $\phi$ obeys \eqref{adler} on a slow time scale.  

The intuitive explanation is that in this limit, the coupling between oscillators is so strong that the population acts like one giant oscillator, with nearly all the microscopic oscillators at the same phase.  Hence the order parameter amplitude remains close to $\rho = 1$  at all times, so the system behaves as if it had a very strongly attracting limit cycle.  This explains why the dynamics of the forced Kuramoto model mimic the Adler equation in this limit.  

For a more analytical route to the same conclusion, look at the large-$K$ behavior of the Takens-Bogdanov point, which essentially lies on the dividing line behind the two big regions.  The formulas~(\ref{TB_Ka})-(\ref{TB_Kb}) imply that 
\begin{equation}
\frac{F_{\rm TB}}{\Omega_{\rm TB}}  \sim  1 - 8 K^{-4}
\end{equation}
 as $K \rightarrow \infty$.  Thus $F \approx \Omega$ for large and even moderate values of $K$.  

\subsection{Comparison to forced van der Pol equation}

For weaker coupling, but still large enough that the system can partially self-synchronize ($2<K<\infty$), the population again acts like a single limit cycle oscillator, but now with a limit cycle that is only weakly attracting.  As before, the complex order parameter plays the role of this effective limit-cycle oscillator.  

So when forcing is applied, we expect the overall dynamics to be like those of a forced, weakly nonlinear oscillator.  And indeed, the stability diagram bears a striking resemblance to that of a forced van der Pol oscillator, in the limit of weak nonlinearity, weak detuning, and weak forcing.  As in the problem studied here, the stability diagram for this well-studied system [Guckenheimer and Holmes 1983] is also organized around a Takens-Bogdanov bifurcation and a saddle-node-loop bifurcation.  

Likewise, some of the regions in the van der Pol diagram are unusually thin and small.  This helps to explain why they were overlooked for decades, until the theory of the Takens-Bogdanov bifurcation was developed and guided later researchers to the missing transitions that, on topological grounds,  had to be there.  

One always expects small regions in systems with Takens-Bogdanov bifurcations because, according to normal form theory, the saddle-node, Hopf, and homoclinic curves have to intersect \emph{tangentially} at the Takens-Bogdanov point.  But here, as in the van der Pol problem, the regions are even smaller still, because they must also hug the line $F \approx \Omega$, for the reasons given above.  

\subsection{Caveats}
It is important to understand what has---and has not---been shown by the analysis presented in this paper.  Following Ott and Antonsen [2008], we made a number of very particular choices in the course of reducing an infinite-dimensional problem to a two-dimensional one.  We chose a special family of initial states (see Eq.\eqref{trick}) and showed that they formed an invariant manifold.  In other words, if the condition \eqref{trick} is satisfied initially, it is automatically satisfied for all time.  Then we chose a special distribution of natural frequencies (see Eq.\eqref{lorentzian}), and required further that the initial state satisfies certain strong analyticity properties with respect to its dependence on these frequencies.  Taken together, these choices then implied that the system's order parameter evolves according to the two-dimensional dynamical system \eqref{orderparamtimeevol}.

If the conclusions that followed were sensitive to these choices, we would not have accomplished much.  But there is reason to believe that the results are robust, and largely independent of these choices.  The strongest evidence is numerical.  Every time we have run simulations of the forced Kuramoto model \eqref{forcedKuramoto_labframe} for hundreds or thousands of oscillators, we have seen all the attractors and bifurcations predicted by the analysis, where they are supposed to be.  Ott and Antonsen [2008] found similar agreement when they studied other variants of the Kuramoto model.  

This suggests that the flow on the invariant manifold faithfully captures the macroscopic dynamics of the full system, at least in some sense.  Unfortunately, we do not know how to make this statement precise.  The issue is probably subtle.  We do not believe, for example, that the invariant manifold is everywhere transversely attracting---it certainly isn't in other problems we have studied. For example, applying the method of Ott and Antonsen [2008] to the Kuramoto model with a bimodal frequency distribution, we found that the invariant manifold in that case could be transversely \emph{repelling} at certain points [Martens et al. 2008]. 

Nor are we sure whether all the attractors for the full system lie within the invariant manifold.  If they did, that would explain why this manifold controls the system's long-term macroscopic dynamics.  But we have no proof of this weaker statement either.

Now regarding the choice of a Lorentzian frequency distribution: this was crucial to the analysis, but not, we suspect, to the results.  Sakaguchi [1988] used a Gaussian $g(\omega)$ and found the same attractors and bifurcations as we did.  Our own simulations for the Gaussian case (unpublished) show that the stability diagram is different in numerical details, of course, but its topology is unaffected.  

On the other hand, the algebraic form of the forced Kuramoto model, with its purely sinusoidal coupling and driving, probably \emph{is} crucial.  The ansatz \eqref{trick} no longer works if the model contains higher harmonics.  Indeed, the bifurcation behavior of the classical (unforced) Kuramoto model is known to be altered when generic periodic functions are used in place of a pure sine function in the coupling [Daido 1994, Crawford 1995].  So we expect new phenomena to arise in the forced Kuramoto model as well, when one departs from pure sinusoids in the driving and coupling terms.

\section{Acknowledgements}

\noindent We thank Ed Ott and Tom Antonsen for sharing their preprint with us.  Our research was supported in part by the National Science Foundation through an NSF Graduate Research Fellowship to L.M.C. and grant DMS-0412757 to S.H.S.

\newpage

\begin{figure}[h!]
\begin{center}
\epsfxsize=6in
\epsffile{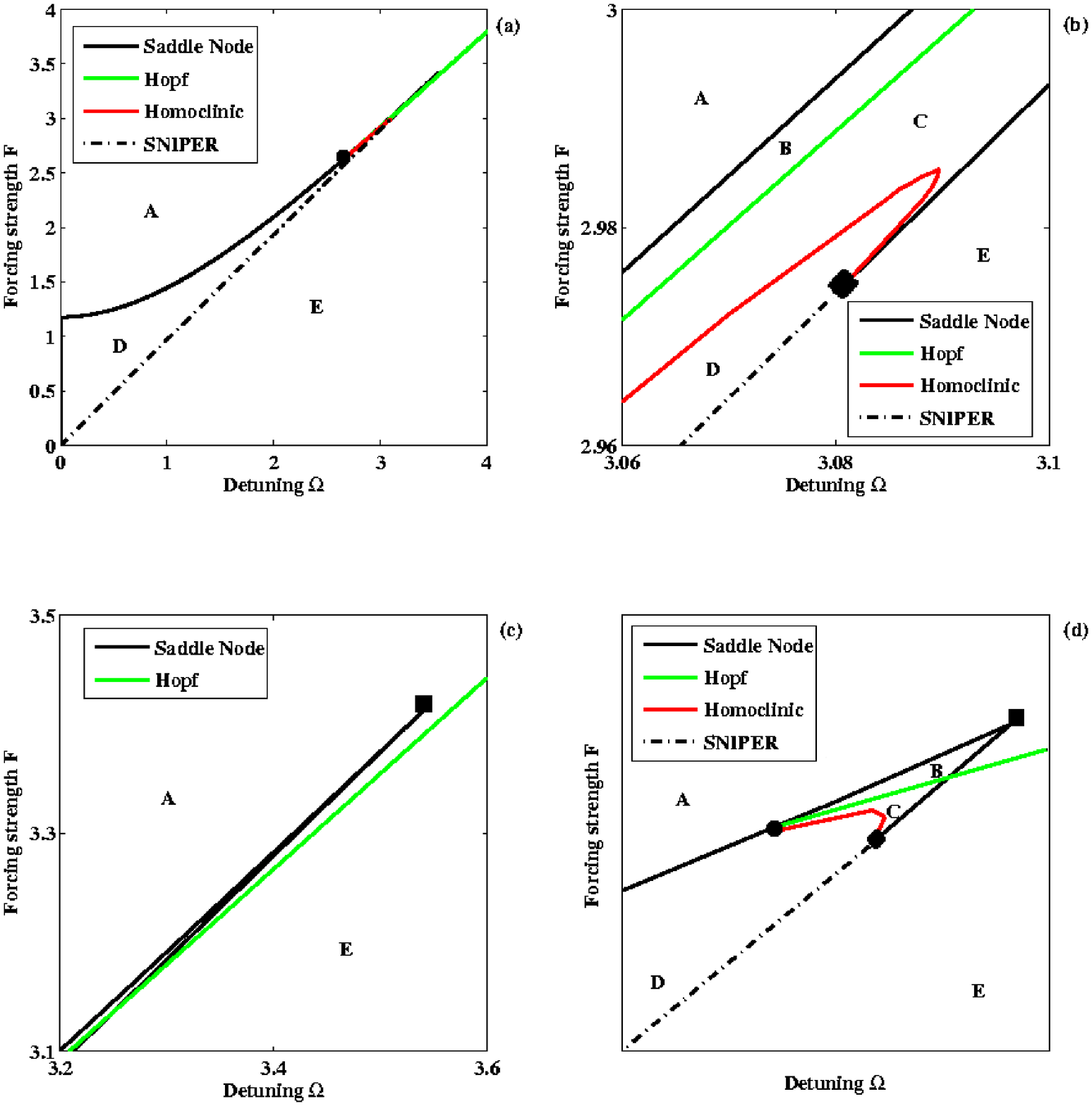}
\end{center}
\end{figure}

\begin{figure}[h!]
\begin{center}
\epsfxsize=4in
\epsffile{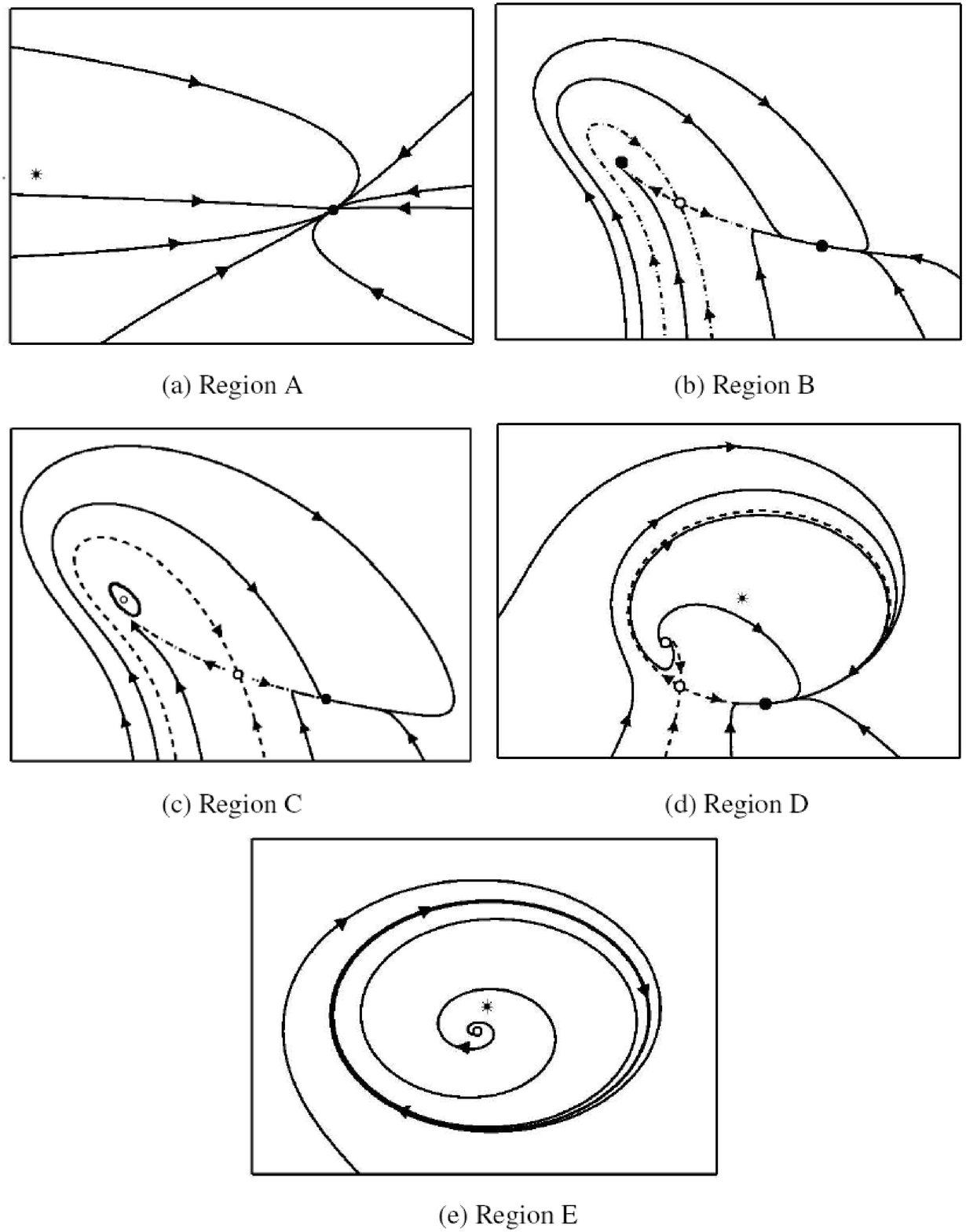}
\end{center}
\end{figure}

\clearpage

\noindent \textbf{Figure Captions}

\bigskip

\textbf{Figure 1:}  Stability diagram for the forced Kuramoto model.  Bifurcation curves are shown with respect to the strength $F$ and detuning $\Omega$ of the external forcing, both of which have been non-dimensionalized by the width $\Delta$ of the distribution of the oscillators' natural frequencies.  The dimensionless coupling strength is fixed at $K=5$. 

(a)  Regions A-E correspond to qualitatively different phase portraits; see text and Fig.~2 for explanations. Four types of bifurcations occur:  supercritical Hopf bifurcation; homoclinic bifurcation; and two types of saddle-node bifurcations.  The saddle-node bifurcations on the upper branch, and also those on the lower branch between the cusp and the saddle-node-loop point,  are purely local.  In contrast, those on the portion of the lower branch extending from the origin to the saddle-node-loop point have global consequences; they are saddle-node infinite-period bifurcations, or SNIPERs, which create or destroy limit cycles.   The filled circle marks a codimension-2 Takens-Bogdanov point, at which the Hopf, homoclinic, and upper saddle-node curve intersect tangentially.  

(b) Enlargement of the cross-over region, just to the right of the Takens-Bogdanov point, where all four bifurcation curves run nearly parallel to one another.  Three of them (Hopf, SNIPER, and the lower branch of saddle-node bifurcations) meet at a codimension-2 saddle-node-loop point, marked by a filled diamond.  

(c)  Enlargement of the region near the codimension-2 cusp point (filled square), where the upper and lower branches of saddle-node bifurcations meet tangentially.  The two branches are almost indistinguishable  in this image.  

(d) Schematic version of the stability diagram, intended to show how the bifurcation curves connect in the confusing cross-over region.  Tangential intersections have been opened up for clarity. 

\bigskip

\textbf{Figure 2: } Phase portraits for the two-dimensional dynamics of the complex order parameter $z$, or equivalently, for the variables $\rho, \phi$ regarded as polar coordinates.  Open dots, unstable fixed points.  Closed dots, stable fixed points. Asterisk, origin of the $z$-plane. Dashed curves, stable and unstable manifolds of the saddle point. The panels are not all shown at the same scale; the regions shown in (b) and (c) are small and have been blown up here for clarity.

\end{document}